\documentclass[journal,twoside,web]{tiistyle}
\usepackage{TII-26-2461.R1}
\usepackage{cite}
\usepackage{amsmath,amssymb,amsfonts}
\usepackage{algorithmic}
\usepackage{graphicx}
\usepackage{textcomp}

\usepackage{amsmath,amsfonts}
\usepackage{algorithmic}
\usepackage{array}
\usepackage{stfloats}
\usepackage{url}
\usepackage{verbatim}
\usepackage{graphicx}
\usepackage{graphicx}
\usepackage[caption=false,font=footnotesize]{subfig}
\usepackage{amsmath}
\usepackage{graphicx}
\usepackage{booktabs} 
\usepackage{adjustbox}
\usepackage{float} 
\usepackage{booktabs}
\usepackage{array}
\usepackage{bm}

\usepackage{booktabs}
\usepackage{tabularx}
\usepackage{array}
\usepackage{makecell}

\usepackage{algorithm}
\usepackage{algorithmic} 

\usepackage{enumerate}
\usepackage{amsmath}

\usepackage{algorithm}
\usepackage{algorithmic}

\usepackage{multirow}
\usepackage{xcolor}

\usepackage{colortbl} 
\usepackage{setspace}
\usepackage{comment}

\makeatletter
\newcommand{\BlueAlgorithmCaption}{%
	\let\old@makecaption\@makecaption
	\long\def\@makecaption##1##2{%
		\old@makecaption{\textcolor{blue}{##1}}{\textcolor{blue}{##2}}%
	}%
}
\newcommand{\EndBlueAlgorithmCaption}{%
	\let\@makecaption\old@makecaption
}
\makeatother

\newcolumntype{Z}{>{\centering\arraybackslash}X} 

\makeatletter
\let\oldbibitem\bibitem

\renewcommand{\bibitem}[2][]{%
	\normalcolor 
	
	\@ifundefined{bibcolor@#2}{}{%
		\color{\csname bibcolor@#2\endcsname}%
	}%
	
	\if\relax\detokenize{#1}\relax
	\oldbibitem{#2}%
	\else
	\oldbibitem[#1]{#2}%
	\fi
}

\newcommand{\colorbib}[2]{%
	\expandafter\def\csname bibcolor@#1\endcsname{#2}%
}
\makeatother

\def\BibTeX{{\rm B\kern-.05em{\sc i\kern-.025em b}\kern-.08em
		T\kern-.1667em\lower.7ex\hbox{E}\kern-.125emX}}
\begin{document}
	\title{Intelligent Domain Adaptation for Power System Transient Stability Assessment Under Varying Operating Scenarios}
	\author{Yuan Yang, Lipeng Zhu, ~\IEEEmembership{Senior Member,~IEEE}, Chao Deng, Jiayong Li, ~\IEEEmembership{Senior Member,~IEEE}, Quan Zhou, ~\IEEEmembership{Senior Member,~IEEE}, and Cong Zhang, ~\IEEEmembership{Senior Member,~IEEE}
		\thanks{This work was supported in part by the National Natural Science Foundation of China under Grants 52207094, 52377095, and 52477090, and in part by the Science and Technology Innovation Program of Hunan Province under Grants 2025RC1023, 2024RC3110, and 2023RC3114. \textit{(Corresponding author: Lipeng Zhu)}}
		\thanks{Yuan Yang is with the National Graduate College for Elite Engineers, Hunan University, Changsha 410082, China (e-mial: yangyuan123654@hnu.edu.cn).}
		\thanks{Lipeng Zhu, Chao Deng, Jiayong Li, Quan Zhou, and Cong Zhang are with the College of Electrical and Information Engineering, Hunan University, Changsha 410082, China (e-mial: lpzhu@hnu.edu.cn; dengchaoll@hnu.edu.cn; jyli@hnu.edu.cn; zquan@hnu.edu.cn; zcong@hnu.edu.cn).}
	}
\maketitle

\begin{abstract}
While deep learning–based transient stability assessment (TSA) approaches have exhibited great potential in power system stability monitoring,  they are prone to undergo performance degradation in practical contexts with frequent variations of operating conditions. To address this issue, this work develops an adaptive  TSA  framework via domain adaptation-enabled deep transfer learning. First, for the sake of capturing the primary transient stability characteristics, a robust metric, i.e., heterogeneous hybrid distribution metric (HHDM), is designed through mathematical means to effectively handle multi-scale Gaussian and long-tail distributions of transient responsive data and to precisely quantify the intrinsic distributional discrepancies between the source and target domains (corresponding to different operating scenarios). With the help of the HHDM, a Bayesian theory-based dual-distribution domain adaptation method is constructed, aligning not only marginal probability distributions between domains but also the distributions of sub-domain categories. Such alignments enable fine-grained transient stability feature transfer, helping significantly improve the adaptability of a well-trained TSA model to target domains. Furthermore, a multilayer sparse regularization algorithm is introduced to mitigate feature volatility caused by variations in operating scenarios, thereby enhancing the model’s generalization in the presence of unforeseen scenarios. Numerical tests on three test systems illustrate that, compared with conventional methods, the proposed framework improves online TSA accuracy by  0.5\%$\sim$5\% in a cost-effective manner, with the learning cost for TSA model update largely reduced.

\end{abstract}
\begin{IEEEkeywords}
 Deep learning, domain adaptation, feature transfer, transient stability assessment.
\end{IEEEkeywords}

\section{Introduction}
\label{sec:introduction}
\IEEEPARstart{W}{ith} the increasing integration of renewable energy sources and power electronic devices, modern power grids are evolving into highly complex and dynamically varying systems. These changes make the major task of maintaining secure and stable system operation more and more challenging, greatly increasing the risk of system transient instability after large disturbances. Therefore, it is essential to develop a real-time, reliable, and accurate transient stability assessment (TSA) scheme to monitor system transient stability status, preventing significant economic losses and severe social impacts caused by losing transient stability \cite{zhu2021networked}.

Traditional TSA approaches, such as time-domain simulation (TDS) and direct methods, are generally based on numerical modeling of the practical systems. However, they rely heavily on precise system modeling and costly computational resources,  making it difficult to meet the efficiency and accuracy requirements of online TSA. Recently, the widespread deployment of phasor measurement units and the rapid development of digital information technologies have significantly improved the capabilities of modern power systems in data collection, storage, and real-time analysis \cite{8326526}. In this context, deep learning (DL)–based methods have become promising alternatives to achieve rapid and accurate TSA for complex power grids \cite{zhu2020intelligent}, \cite{11488592}. Basically, they leverage DL techniques to discover the underlying relationships between initial system states/responses and eventual system stability status from a stability knowledge base prepared offline, whereby rapid TSA can be realized during online application \cite{zhu2021semi}.

There exists an implicit assumption in DL-enabled studies, i.e., the data distributions in offline training and online application are expected to remain consistent \cite{10970109}. In practice, however, this assumption could become invalid since the system topologies, operating points, and transient contingencies involved in online application stages may largely differ from those considered in the offline TSA model training stage \cite{10141875}. Such differences lead to distributional discrepancies between offline learning data and online measured data, being likely to degrade TSA model performance and even induce erroneous TSA results. Typically, this issue can be addressed by extensively gathering new cases under the new operating scenarios and retraining the TSA model from scratch. This seems to be a straightforward solution, yet it is usually time-consuming and requires the collection of large amounts of labeled TSA data in new operating scenarios \cite{9957101}. In practical contexts, it is costly to acquire a large-sized TSA dataset that covers all potential new operating conditions, posing a considerable challenge to continuous TSA model retraining and updating.

Domain adaptation (DA) has emerged as a cost-effective solution to address the above issue, helping achieve strong generalization capability and robustness in new operating scenarios \cite{10135140}. Recently, a number of inspiring DA-based research efforts have emerged across various industrial informatics areas, including fault diagnosis, structural health monitoring, and industrial process monitoring, demonstrating the feasibility and potential of achieving satisfactory performance in new scenarios using only a limited number of target-domain samples. Among the diverse methodologies employed, DA frameworks based on Maximum Mean Discrepancy (MMD) have been widely adopted due to their simplicity and effectiveness. However, these methods seem to share two significant limitations in TSA. First, as illustrated in the sequel of this paper, post-disturbance transient stability data exhibit mixed distributions—consisting of both Gaussian and long-tail patterns. In particular, the latter pattern is characterized by long tails and frequent outliers, which correspond to anomalous values occurring more often than expected due to sudden and severe system disturbances. This violates the default Gaussian assumption underlying the MMD’s Gaussian kernel function, having a high possibility of decreasing its robustness when applied to the TSA task. Second, existing MMD-based studies primarily align the marginal probability distributions of the source and target domains, with no attention to fine-grained alignment at the sub-domain (i.e., class-conditional) level. This omission may result in insufficient domain confusion and suboptimal knowledge transfer, thereby limiting the model’s ability to accurately distinguish stability status under diverse unforeseen operating conditions.

To tackle the above research gap, this work proposes a Heterogeneous Hybrid Distribution Metric-based Adaptive Transient Stability Assessment (HHDM-ATSA) framework. Within the framework, to address the limited robustness of existing DA methods to the long-tail patterns of transient data, a Student kernel is integrated with the conventional Gaussian kernel to devise a novel heterogeneous hybrid distribution metric (HHDM), being capable of accurately characterizing intricate distributional differences between source and target domains under different transient operating scenarios. Based on the well-designed HHDM, a dual-distribution domain adaptation (DDA) method is further developed. The DDA method not only aligns the marginal probability distributions across different domains but also leverages the Bayes theory to realize sub-domain–level distributions alignment, thereby facilitating fine-grained and transferable transient stability feature learning. Moreover, a multilayer sparse regularization (MSR) strategy is devised to obtain a more uniform distribution of transient stability features, enhancing their robustness to distribution shifts and improving representation capability. By doing so, the obtained HHDM-ATSA framework is able to augment a well-trained TSA model’s adaptability across diverse operating conditions, thus contributing to more reliable online TSA in practical power grids. The major contributions and merits of this work are summarized as follows.

\begin{enumerate}
	\item This work develops an intelligent HHDM-ATSA framework, which realizes reliable and robust TSA against diverse and varying online operating scenarios via cost-effective transient stability knowledge transfer. It largely reduces the time cost of TSA model updates without sacrificing online TSA accuracy, which remains a crucial yet unsolved issue in the area of data-driven TSA.
	
	\item A novel distribution metric called HHDM is properly designed to address the long-tail distribution characteristics of transient stability data. Via mathematical optimization techniques, the HHDM precisely captures the intricate distributional differences between various operating scenarios. Compared with conventional distance metrics, this new metric helps build more robust and reliable DA strategies, being more applicable in practical power systems with diverse complicated system dynamics.
	
	\item An MSR technique is carefully devised to encourage more uniform feature learning during TSA model training, helping mitigate the volatility of transient features caused by variations in grid operating conditions. Meanwhile, a DDA method is put forward to align both global and local distributions across domains, contributing to the reduction of the classification complexity of the TSA model. The combination of the MSR and DDA technique enables rapid updates of the TSA model under diverse scenarios, thereby helping achieve highly reliable TSA during continuous online monitoring.
\end{enumerate}

The rest of this paper is organized as follows. Section II first provides an overview of related work in the literature. The basic problems with regard to data-driven TSA are described in Section III. Section IV details the HHDM-ATSA framework to tackle these problems. The implementation of adaptive TSA in Section V. Section VI presents numerical test results for verification. Conclusions are finally drawn in Section VII.

\section{Related Work}
\subsection{Adversarial Training-Based DA}
Adversarial training-based DA methods typically introduce a domain discriminator to align cross-domain distributions,  in which a feature extractor is trained to learn domain-invariant features via a minimax game. In \cite{9099635}, an adversarial unlabeled domain-adversarial transfer network is developed, incorporating a contrast estimation term that quantifies the similarity of fault data distributions via adversarial training, thereby improving cross-domain fault diagnosis performance. Ref.\cite{9247269} designs a Wasserstein distance-based domain-adversarial network that extracts domain-invariant features directly from raw signals via adversarial training. It proves effective in improving the generalization capability of fault diagnosis models across domains, even when labeled data is scarce.

Besides aligning marginal probability distributions, several approaches also further align conditional probability distributions \cite{11134056}. Aiming for more fine-grained DA, \cite{10017183} introduces a domain-conditioned joint adaptation strategy, which learns domain-shared features via domain-adversarial training, with class-level adaptation for extracting class-discriminative features through bi-classifier adversarial training. In \cite{LU2025104172}, a dynamic hybrid DA method is developed to adaptively adjust class-wise distributions during joint alignment, facilitating fine-grained DA.  However, this learning process may suffer from instability and may encounter difficulties in converging to the Nash equilibrium.

\subsection{Statistical Distance-Based DA}
Statistical distance–based DA approaches achieve domain alignment by minimizing explicit metrics between feature distributions, leading to faster training convergence and more reliable predictive performance \cite{10934148}. Classical distance metrics include correlation alignment (CORAL), Wasserstein distance (WD), and MMD \cite{9931460}. Compared with CORAL and WD, MMD provides a lightweight, nonparametric discrepancy metric. It can be easily incorporated into DL and is capable of extracting more general distribution shifts via mean embeddings.

In many areas of industrial informatics, MMD has been extensively applied to quantify distributional differences between domains. Based on MMD, \cite{8058000} designs a DA algorithm for efficient fault diagnosis with limited labeled data. Analogously, to formally mitigate the effects of visual discrepancy, \cite{8454781} develops a framework for deep adaptation networks that extends neural networks to DA problems. In \cite{xiang2023novel}, a deep multi-layer CNN is adopted to extract features from both source and target domains, and learning of domain-invariant features by minimizing the MMD in rotor crack fault diagnosis.

The aforementioned approaches quantify distribution discrepancy at the global level while neglecting class-specific information, which may lead to misalignment and degrade performance \cite{9802910}. Ref.\cite{9745095, 10251665} put forward a joint distribution adaptation approach that extends MMD to measure discrepancies in both global and local distributions across domains, enabling the extraction of appropriate feature representations. To further enhance discriminative feature learning within the DA framework, \cite{rezaei2024visual} learns domain-invariant features by pulling cases from the same class closer while pushing cases from different classes farther apart, thereby yielding more discriminative latent feature representations. However, they merely approximate the conditional probability distribution with the class-conditional probability distribution, introducing approximation error that may, in turn, compromise domain confusion performance. Different from them, the proposed DDA method leverages Bayes’ theorem to explicitly align the true conditional probability distributions across domains, enabling a more fine-grained DA.

\subsection{Relevant Applications in Power Engineering}
In the specific domain of power engineering that covers the subject of power grid TSA in this paper, the abovementioned DA algorithms have also been successfully adopted to address domain shift issues. A brief overview of relevant applications published in recent years is provided below, highlighting the key aspects that distinguish them from this work.

In \cite{li2022adaptive}, an MMD-based adaptive updating mechanism is proposed for TSA. It can flexibly select the optimal transfer paths based on the distributional difference between domains, thus reducing the computational burden of TSA model updating. An improved MMD metric is developed in \cite{yang2024transfer} to quantify distributional discrepancies between actual and presumed faults, effectively improving the model’s accuracy and robustness across various fault scenarios. Furthermore, a hybrid approach combining adversarial training and MMD-based quantification is introduced in \cite{8871201}, yielding improved classification performance under various operating conditions. Overall, the MMD-based DA algorithms involved in these studies \cite{li2022adaptive,yang2024transfer,8871201} are different from the HHDM-ATSA framework in this paper. Specifically, these MMD-based DA approaches do not explicitly account for the long-tail nature of post-disturbance transient stability features, whereas the designed HHDM metric is able to effectively model and handle such long-tail distributions to enhance robustness in TSA.

To sum up, this paper systematically integrates the proposed HHDM metric and the DDA method to construct an intelligent framework for adaptive TSA. Such a unique learning framework and the resulting strong adaptability in adapting to diverse new online operating conditions make this paper different from existing research efforts in the literature.

\section{Problem Description}

Given a specific power system, a DL-based TSA model can be derived by learning the following mapping from $n_s$ transient cases:
\begin{equation}
	\mathcal{F} : \boldsymbol{x}_i \rightarrow y_i, \ 
	\text{for } \boldsymbol{x}_i = \{\boldsymbol{z}_1, \boldsymbol{z}_2, \ldots, \boldsymbol{z}_{n_c}\}, \ 
	y_i \in \{0,1\}
\end{equation}

Here \(\boldsymbol{x}_i \in \mathbb{R}^{n_c \times L}\) denotes the feature matrix of case $i$ (\(1 \leq i \leq n\)), being composed of $n_c$ time series (with \(L\) data points in total) describing trajectory cluster-based features. Taking the $k$th (\(1 \leq k \leq n_c\)) time series for instance, it characterizes the clustered geometric characteristics of all the bus voltage trajectories spanning the pre-fault, fault-on, and post-fault process of case $i$. \(y_i\) represents the stability status of the system \((0 \rightarrow \textit{stable}, \ 1 \rightarrow \textit{unstable})\). The effectiveness of (1) relies on the assumption that the input data of all the transient cases, i.e., \(\boldsymbol{x}\), always reside in the same feature space, which is rarely satisfied in practice due to frequent variations in system operating conditions. Specifically, let $\boldsymbol{\mathcal{D}}_s = \{\boldsymbol{x}_i^s, {y}_i^s\}_{i=1}^{n}$ denote a collection of ${n}$ cases acquired from the historical operating scenarios (source domain) and $\boldsymbol{\mathcal{D}}_t= \{\boldsymbol{x}_j^t, {y}_j^t\}_{j=1}^{m}$ represent a collection of ${m}$ cases acquired from new scenarios (target domain) that differ substantially from $\boldsymbol{\mathcal{D}}_s$. Here, ${y}_i^s$ and ${y}_j^t$ are the corresponding labels of $\boldsymbol{x}_i^s$ and $\boldsymbol{x}_j^t$.  The primary discrepancies between the two domains arise from variations in operating conditions, such as load fluctuations, generator dispatch, network structural changes, etc. Such discrepancies may invalidate the mathematical mapping in (1) learned by a well-trained TSA model, making it difficult for the model to adapt to new operating conditions.

To address this issue, MMD-based DA methods have been explored. The core principle of such methods is to employ MMD to quantify the discrepancy across domains. During TSA model training, distribution alignment is realized by minimizing this distance, thereby enhancing the model's adaptability to new scenarios. Let $\mathcal{H}$ denote the reproducing kernel Hilbert space, the MMD between $\boldsymbol{\mathcal{D}}_s$ and $\boldsymbol{\mathcal{D}}_t$ is defined as:

\begin{equation}
	\scalebox{0.825}{$
		\begin{aligned}
			\mathrm{MMD}[\mathcal{H}, \boldsymbol{x}^{s}, \boldsymbol{x}^{t}] &= \sup_{\| h \|_{\mathcal{H}} \leq 1} \left\langle h, \mathbb{E}_{\boldsymbol{x}^{s} \sim p} k_\sigma^g(\boldsymbol{x}^{s}, \cdot) - \mathbb{E}_{\boldsymbol{x}^{t} \sim q} k_\sigma^g(\boldsymbol{x}^{t}, \cdot) \right\rangle_\mathcal{H} \\
			&= \|h\|_\mathcal{H} \left\| \mathbb{E}_{\boldsymbol{x}^{s} \sim p} k_\sigma^g(\boldsymbol{x}^{s}, \cdot) - \mathbb{E}_{\boldsymbol{x}^{t} \sim q} k_\sigma^g(\boldsymbol{x}^{t}, \cdot) \right\|_\mathcal{H} \\
			&= \left\| \mathbb{E}_{\boldsymbol{x}^{s} \sim p} k_\sigma^g(\boldsymbol{x}^{s}, \cdot) - \mathbb{E}_{\boldsymbol{x}^{t} \sim q} k_\sigma^g(\boldsymbol{x}^{t}, \cdot) \right\|_\mathcal{H} \\
			&\approx \left\| \frac{1}{n} \sum_{i=1}^{n} \phi(\boldsymbol{x}_i^s) - \frac{1}{m} \sum_{j=1}^{m} \phi(\boldsymbol{x}_j^t) \right\|_{\mathcal{H}}
		\end{aligned}
		$}
\end{equation}
where $p$ and $q$ are the underlying true distributions of  $\boldsymbol{\mathcal{D}}_s$ and $\boldsymbol{\mathcal{D}}_t$, $\mathbb{E}$ denotes the expectation operator, $h$ is a nonlinear continuous function in $\mathcal{H}$, and $\| h \|_{\mathcal{H}} \leq 1$ ensures the existence of an upper bound, $\langle \cdot,\cdot \rangle_\mathcal{H}$ represents the inner product in $\mathcal{H}$, \( k_\sigma^g(*, \cdot) \) is the Gaussian kernel function with the corresponding feature mapping \( \phi(\cdot) \), i.e., \( k_\sigma^g(*, \cdot) = \langle \phi(*), \phi(\cdot) \rangle_{\mathcal{H}} \). The kernel $k_\sigma^g(\boldsymbol{x}^{s}, \boldsymbol{x}^{t})$ is defined as:

\begin{equation}
	k_\sigma^g(\boldsymbol{x}^{s}, \boldsymbol{x}^{t})= \exp\left( -\frac{\|\boldsymbol{x}^{s} - \boldsymbol{x}^{t}\|^2}{2\sigma^2} \right)
\end{equation}
where $\sigma$ is the width of the kernel function,  $\|\cdot\|^2$ denotes the squared Euclidean distance, and $\exp(\cdot)$ represents the exponential function.

\begin{figure}[t]
	\centering
	\includegraphics[width=0.33\textwidth]{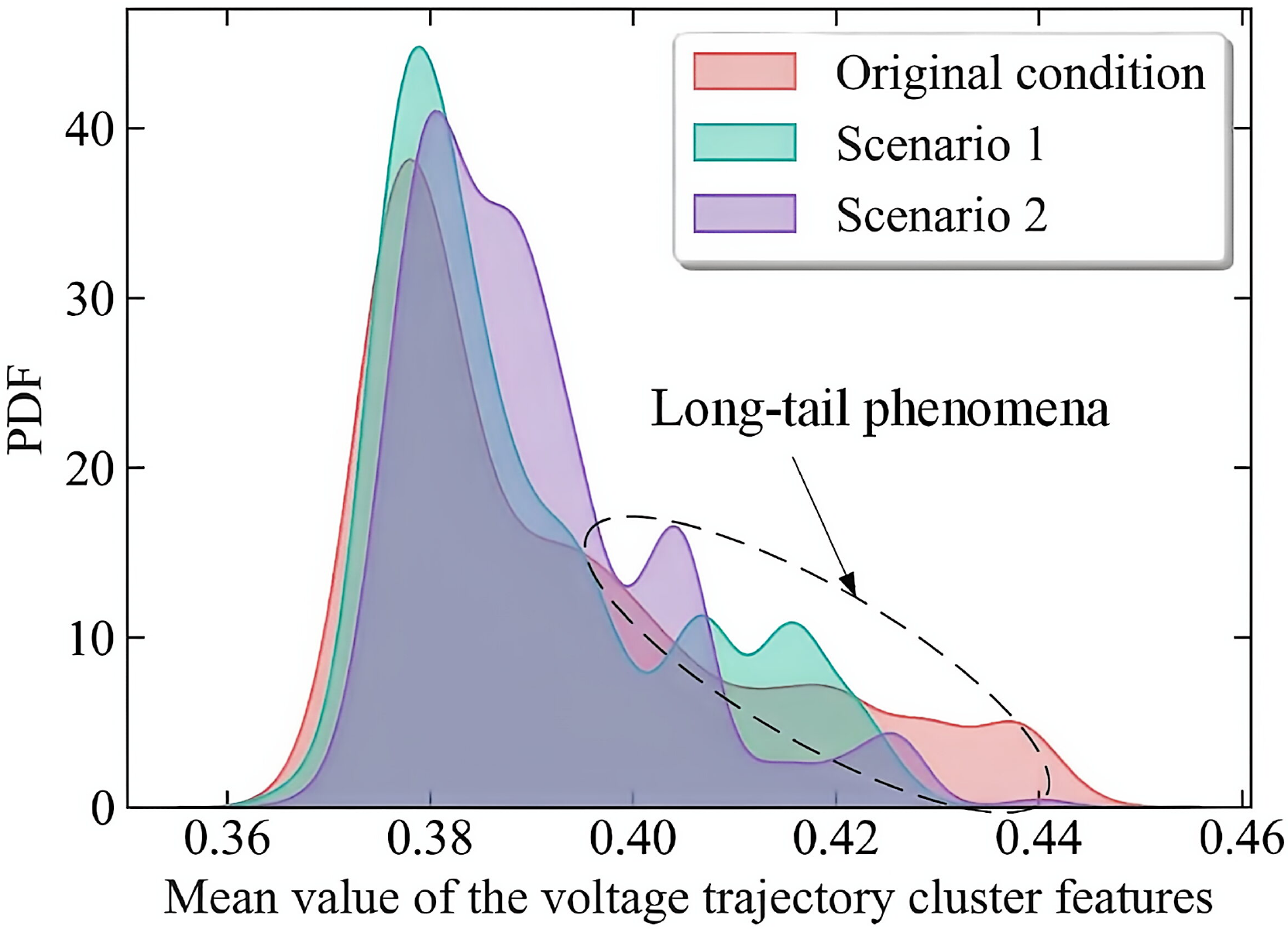}
	\setlength{\abovecaptionskip}{-5pt}
	\caption{PDFs of the mean values of voltage trajectory cluster features.}
	\label{fig:probability density}
\end{figure}

However, power systems typically exhibit nonlinear and complex dynamics under large disturbances. Taking the original and new conditions of the IEEE 39-bus benchmark system as an illustrative example, Fig. 1 depicts the probability density functions (PDFs) of the mean value \(m_v\) of the clustered voltage trajectory features obtained from the pre-fault, fault-on, and post-fault stages of the transient event process caused by a transient fault. Both the original and new conditions exhibit Gaussian distributions on the left side, but pronounced long tails on the right side, indicating a higher probability of the presence of extreme values (i.e., \(m_v \geq 0.4\), with the occurrence probability being considerably higher than expected) in system transient responsive data. These mixed Gaussian and long-tailed characteristics undermine the robustness of the Gaussian kernel in handling transient response data, which are originally suitable for handling light-tailed data. Moreover, evident distributional shifts between the original and new scenarios are observed. These facts make it invalid to apply the mapping in (1) represented by a specific TSA model to new operating conditions, thus degrading the reliability of the TSA model in practical contexts with varying operating conditions.

\section{Proposed HHDM-ATSA FRAMEWORK}
The proposed HHDM-ATSA framework is shown in Fig. 2. It first takes two parameter-sharing feature extractors to capture transient stability features with strong cross-domain capability. For convenience, the simple yet efficient one-dimensional convolutional neural network (1D-CNN) is taken as the primary algorithm to build the two feature extractors. Note that the proposed framework is not limited to implementing TSA via 1D-CNN. Instead, it can be compatibly configured with other DL algorithms for feature learning. Next, the HHDM is designed to explicitly model the long-tailed patterns of transient stability data, thereby enabling the accurate quantification of the mean embedding distance between transient stability data distributions. Based on the HHDM, the DDA strategy quantifies the discrepancies between the source and target domains from both global and local perspectives (corresponding to marginal and conditional probability distributions). The model parameters are then iteratively optimized to minimize these discrepancies, thereby facilitating fine-grained domain alignment and consequently implementing efficient and precise stability prediction in new scenarios. More details about the key techniques are presented in the following sections.


\begin{figure*}[t]
	\centering
	\includegraphics[width=0.79\textwidth]{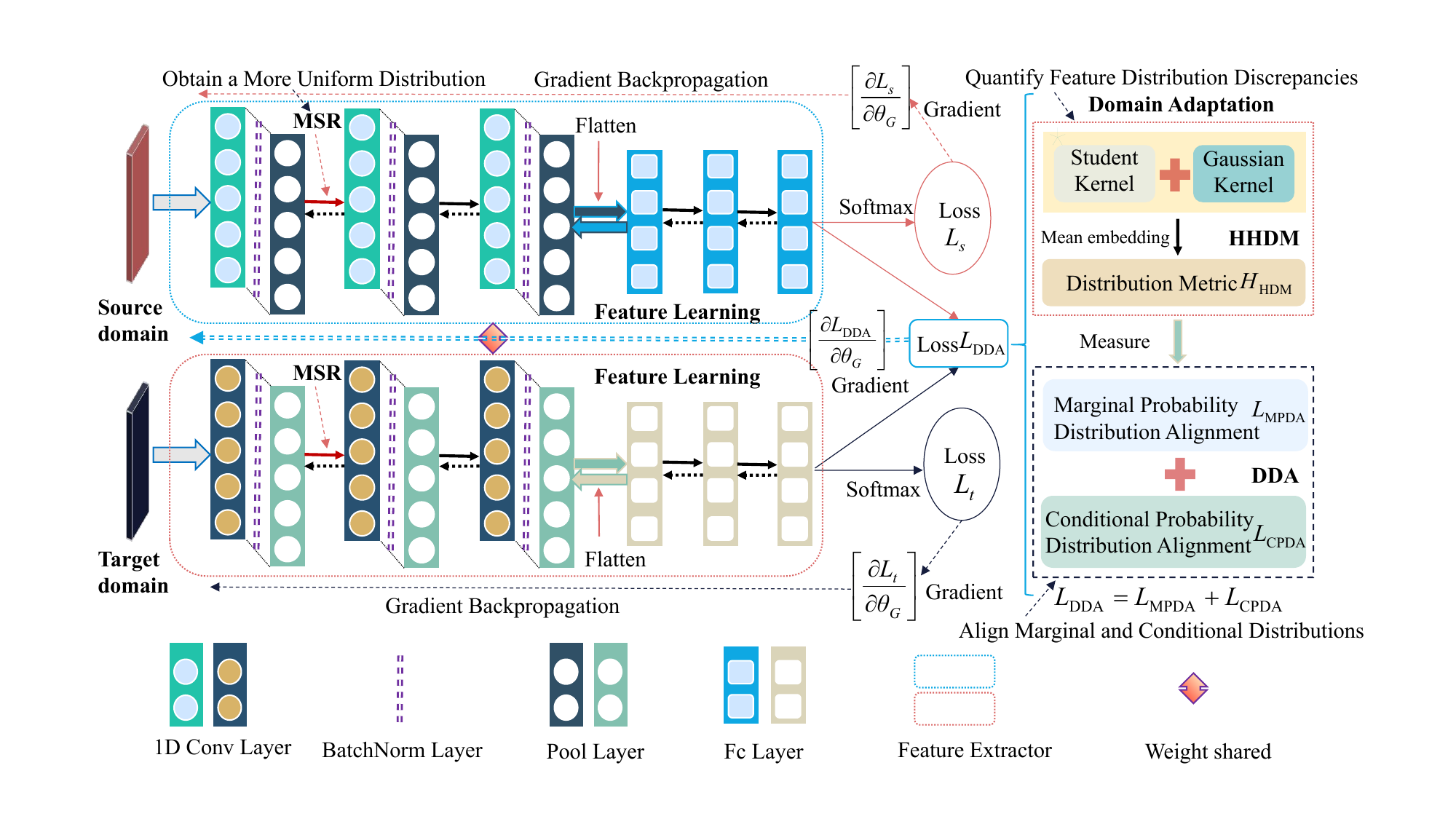}
	
	\caption{Proposed HHDM-ATSA framework.}
	\label{fig:ATSA-HHDM framework}
\end{figure*}

\subsection{Heterogeneous Hybrid Distribution Metric}

\subsubsection{Heterogeneous Hybrid Kernel Function}
As mentioned in Section II, the Gaussian kernel function exhibits limited robustness in modeling transient stability data with pronounced long-tail behaviors. In contrast, the Student-\(t\) distribution, due to its long-tailed property, demonstrates stronger capability in capturing data patterns with extreme values. To jointly accommodate both Gaussian and long-tail features inherent in transient stability data, this work proposes a novel heterogeneous hybrid kernel function:
\begin{equation}
	h_{\text{gs}}(\boldsymbol{x}^{s}, \boldsymbol{x}^{t}) = \alpha \sum_{i=1}^{n} \eta_i k_{\sigma_i}^g(\boldsymbol{x}^{s}, \boldsymbol{x}^{t}) + \beta \sum_{i=1}^{m} \mu_i k_{d_i}^t(\boldsymbol{x}^{s}, \boldsymbol{x}^{t})
\end{equation}
where 
\( \alpha + \beta = 1 \), 
\( \sum_{i=1}^{n} \eta_i = 1 \), 
\( \sum_{j=1}^{m} \mu_j = 1 \), 
and \( k_{d_i}^t(\boldsymbol{x}^{s},\boldsymbol{x}^{t}) \) denotes the Student kernel with degrees of freedom \( d_i \), 
derived from the Student-\(t\) distribution as:
\begin{equation}
	k_d^t(\boldsymbol{x}^{s}, \boldsymbol{x}^{t}) = \left(1 + \frac{\|\boldsymbol{x}^{s} - \boldsymbol{x}^{t}\|^2}{d} \right)^{-\frac{d+1}{2}}
\end{equation}

The validity of (4) and (5) as kernel functions can be demonstrated using Mercer’s condition together with the fundamental properties of kernels.

In summary, \( h_{\text{gs}}(\boldsymbol{x}^{s}, \boldsymbol{x}^{t}) \) effectively handles both multi-scale Gaussian and long-tail distributions within transient stability data, thus improving the DA framework’s robustness and adaptability under diverse operating conditions in grids.

\subsubsection{HHDM-Based Distributional Discrepancy Measurement}
For reliable cross-domain distribution alignment, it is crucial to accurately quantify the intrinsic discrepancies between $\boldsymbol{\mathcal{D}}_s$ and $\boldsymbol{\mathcal{D}}_t$. To this end, the HHDM is designed on the basis of \( h_{\text{gs}}(\boldsymbol{x}^{s}, \boldsymbol{x}^{t}) \) to quantify the mean embedding distance between the source and target domains, thereby offering a rigorous characterization of their overall transient distributional discrepancies, is now formulated as:
\begin{equation}
	\begin{aligned}
		& H_{\text{HDM}}\left[\mathcal{H}', \boldsymbol{x}^{s}, \boldsymbol{x}^{t}\right]\\
		&= \sup_{\|h\|_{\mathcal{H}'} \leq 1} \left\langle h, \mathbb{E}_{\boldsymbol{x}^{s} \sim p} \, h_{\text{gs}}(\boldsymbol{x}^{t}, \cdot) - \mathbb{E}_{\boldsymbol{x}^{t} \sim q} \, h_{\text{gs}}(\boldsymbol{x}^{t}, \cdot) \right\rangle_{\mathcal{H}'} \\
		&= \left\| h \right\|_{\mathcal{H}'} \left\| \mathbb{E}_{\boldsymbol{x}^{s} \sim p} \, h_{\text{gs}}(\boldsymbol{x}^{s}, \cdot) - \mathbb{E}_{\boldsymbol{x}^{t} \sim q} \, h_{\text{gs}}(\boldsymbol{x}^{t}, \cdot) \right\|_{\mathcal{H}'} \\
		&= \left\| \mathbb{E}_{\boldsymbol{x}^{s} \sim p} \, h_{\text{gs}}(\boldsymbol{x}^{s}, \cdot) - \mathbb{E}_{\boldsymbol{x}^{t} \sim q} \, h_{\text{gs}}(\boldsymbol{x}^{t}, \cdot) \right\|_{\mathcal{H}'}\\
		&\approx \left\| \frac{1}{n} \sum_{i=1}^{n} \phi'(\boldsymbol{x}_i^s) - \frac{1}{m} \sum_{j=1}^{m} \phi'(\boldsymbol{x}_j^t) \right\|_{\mathcal{H}'}
	\end{aligned}
\end{equation}
where $\mathcal{H}'$ denotes the multi-group Hilbert space defined by \( h_{\text{gs}}(\boldsymbol{x}^{s}, \boldsymbol{x}^{t}) \), \( m \) and \( n \) are the number of cases in $\boldsymbol{\mathcal{D}}_s$ and $\boldsymbol{\mathcal{D}}_t$, and  $\phi'(\cdot)$ is the feature mapping associated with \( h_{\text{gs}}(\boldsymbol{x}^{s}, \boldsymbol{x}^{t}) \).

For the convenience of the calculation of the kernel-based expectation form, the square operation is applied to HHDM:
\begin{equation}
	H_{\text{HDM}}^2\left[\mathcal{H}', \boldsymbol{\mathcal{D}}_s, \boldsymbol{\mathcal{D}}_t\right] =
	\left\| \mathbb{E}_{\boldsymbol{x}^{s} \sim p} \, h_{\text{gs}}(\boldsymbol{x}^{s}, \cdot) - 
	\mathbb{E}_{\boldsymbol{x}^{t} \sim q} \, h_{\text{gs}}(\boldsymbol{x}^{t}, \cdot) \right\|_{\mathcal{H}'}^2
\end{equation}

Therefore, the biased empirical statistic of \( H_{\text{HDM}}^2 \), denoted by \( H_{\text{HDM}_b}^2 \), is defined as:
\begin{equation}
	\begin{aligned}
		H_{\text{HDM}_b}^2\left[\mathcal{H}', \boldsymbol{\mathcal{D}}_s, \boldsymbol{\mathcal{D}}_t\right] 
		= & \; \frac{1}{m^2} \sum_{i=1}^{m} \sum_{j=1}^{m} h_{\text{gs}}(\boldsymbol{x}_{i}^s, \boldsymbol{x}_{j}^s) \\
		+ \frac{1}{n^2} \sum_{i=1}^{n} \sum_{j=1}^{n} h_{\text{gs}}(\boldsymbol{x}_{i}^t, \boldsymbol{x}_{j}^t) 
		& - \frac{2}{mn} \sum_{i=1}^{m} \sum_{j=1}^{n} h_{\text{gs}}(\boldsymbol{x}_{i}^s, \boldsymbol{x}_{j}^t)
	\end{aligned}
\end{equation}

Analogously, the unbiased empirical statistic of \( H_{\text{HDM}}^2 \), denoted as \( H_{\text{HDM}_u}^2 \), can be expressed as:
\begin{equation}
	\hspace*{-1.2cm}
	\begin{aligned}
		H_{\text{HDM}_u}^2\left[\mathcal{H}', \boldsymbol{\mathcal{D}}_s, \boldsymbol{\mathcal{D}}_t\right] 
		= \frac{1}{m(m-1)} \sum_{i=1}^{m} \sum_{\substack{ j \neq i}}^{m} h_{\text{gs}}(\boldsymbol{x}_{i}^s, \boldsymbol{x}_{j}^s) \\
		+ \frac{1}{n(n-1)} \sum_{i=1}^{n} \sum_{\substack{j \neq i}}^{n} h_{\text{gs}}(\boldsymbol{x}_{i}^t, \boldsymbol{x}_{j}^t) - \frac{2}{mn} \sum_{i=1}^{m} \sum_{j=1}^{n} h_{\text{gs}}(\boldsymbol{x}_{i}^s, \boldsymbol{x}_{j}^t)
	\end{aligned}
\end{equation}

For \(H_{\text{HDM}_b}^2 \) and \( H_{\text{HDM}_u}^2 \), if sufficient cases are available in $\boldsymbol{\mathcal{D}}_t$, \(H_{\text{HDM}_b}^2 \) is preferred, as it leverages the entire dataset to yield a more stable estimate of cross-domain discrepancies. Otherwise, in the context of case scarcity, \( H_{\text{HDM}_u}^2 \) is favored due to its capability of mitigating empirical bias and reducing estimation errors arising from limited data.

\subsection{Dual-Distribution Domain Adaptation}

For the TSA task, variations of system operating conditions may induce distribution shifts, such that \( P(\boldsymbol{x}^{s}) \neq P(\boldsymbol{x}^{t}) \) or \( P(\boldsymbol{y}^{s} \mid \boldsymbol{x}^{s}) \neq P(\boldsymbol{y}^{t} \mid \boldsymbol{x}^{t}) \). To address this issue, a DDA strategy is developed, where HHDM is leveraged to construct a loss function that jointly quantifies the discrepancies of both marginal and conditional probability distributions across domains. By iteratively minimizing these discrepancies, the feature extractor captures transient domain-invariant features, thus enhancing the model’s adaptability to new operating conditions. The underlying mechanism is detailed as follows.

First, since the marginal probability distribution directly reflects the global characteristics of cases \cite{5640675}, a marginal probability distribution alignment (MPDA) mechanism is introduced. The associated loss function $L_{\mathrm{MPDA}}$ is described~by
\begin{equation}
	L_{\mathrm{MPDA}} = \left\| \mathbb{E}_{P(\boldsymbol{x}^{s})} \left[ G(\boldsymbol{x}^{s}) \right] - \mathbb{E}_{P(\boldsymbol{x}^{t})} \left[ G(\boldsymbol{x}^{t}) \right] \right\|_{\mathcal{H}_{\text{HDM}^*}^{2}}
\end{equation}
where $G(\cdot)$ denotes two parameter-sharing feature extractors.

However, marginal probability distribution alignment alone may cause same-class cases from different domains to be misaligned, yielding ambiguous decision boundaries. Considering the intrinsic correlations among same-class cases across domains, it is highly desirable to further perform alignment at the subdomain level. To this end, by leveraging Bayes’ theorem, the conditional probability can be reformulated into a class-conditional form \cite{qian2023deep}, expressed as:
\begin{equation}
	P(\boldsymbol{y}^{s} = c \mid \boldsymbol{x}^{s}) = \frac{P(\boldsymbol{y}^{s} = c) \, P(\boldsymbol{x}^{s} \mid \boldsymbol{y}^{s} = c)}{P(\boldsymbol{x}^{s})}
\end{equation}
\begin{equation}
	P(\boldsymbol{y}^{t} = c \mid \boldsymbol{x}^{t}) = \frac{P(\boldsymbol{y}^{t} = c) \, P(\boldsymbol{x}^{t} \mid \boldsymbol{y}^{t} = c)}{P(\boldsymbol{x}^{t})}
\end{equation}
where $P(\boldsymbol{x}^{s} \mid \boldsymbol{y}^{s} = c)$ and $P(\boldsymbol{x}^{t} \mid \boldsymbol{y}^{t} = c)$ denote the class-conditional probabilities for class \( c \) in the source and target domains, respectively, reflecting the confidence of the TSA model in classifying cases. \( P(\boldsymbol{y}^{s} = c) \) and \( P(\boldsymbol{y}^{t} = c) \) denote the prior probabilities of class \( c \) in the two domains, defined~as:
\begin{equation}
	P(\boldsymbol{y}^{\mathrm{s}}=c)=\frac{n^{c}}{\sum_{i=1}^{R} n^{i}},\quad
	P(\boldsymbol{y}^{\mathrm{t}}=c)=\frac{m^{c}}{\sum_{i=1}^{R} m^{i}}
\end{equation}
where \( R \) denotes the number of classes,
\( n^c \) and \( m^c \) represent the number of cases belonging to class \( c \) 
in the $\boldsymbol{\mathcal{D}}_s$ and $\boldsymbol{\mathcal{D}}_t$, respectively. That is,$\sum_{c=1}^{R} n^c = n$ and $\sum_{c=1}^{R} m^c = m$.

By combining (11) and (12), a conditional probability distribution alignment (CPDA) mechanism is designed. 
Its loss function \( L_{\text{CPDA}} \) is defined as:
\begin{align}
	\hspace*{-1.4em}
	L_{\text{CPDA}} &= \varphi(\varepsilon) \sum_{c=1}^{R} \left\| \mathbb{E}_{P(\boldsymbol{x}^s|\boldsymbol{y}^s=c)} \left[ G(\boldsymbol{x}^s) |\boldsymbol{y}^s=c \right] P(\boldsymbol{y}^{s}=c) \right. \notag \\
	&\quad - \left. \mathbb{E}_{P(\boldsymbol{x}^t|\boldsymbol{y}^t=c)} \left[ G(\boldsymbol{x}^{t}) |\boldsymbol{y}^t=c \right] P(\boldsymbol{y}^{t}=c) \right\|_{\mathcal{H}_{\text{HDM}^*}^{2}}
\end{align}
where \( \varphi(\varepsilon) = \left( \frac{2}{1 + \exp(-\chi \varepsilon)} - 1 \right) \), \( \chi = 10 \) controls the growth rate of the weighting function, and \( \varepsilon \) denotes the ratio of the current training iteration to the total number of iterations.

\( L_{\mathrm{CPDA}} \) quantifies the discrepancies between the conditional probability distributions of the $\boldsymbol{\mathcal{D}}_s$ and $\boldsymbol{\mathcal{D}}_t$, thereby guiding the alignment of same-class cases across domains.

By integrating the MPDA and CPDA mechanisms, the DDA loss function \( L_{\text{DDA}} \) is defined as:
\begin{equation}
	\scalebox{0.91}{$
		\begin{aligned}
			L_{\text{DDA}} 
			= & \left\| \mathbb{E}_{P(\boldsymbol{x}^{s})}[G(\boldsymbol{x}^{s})] - \mathbb{E}_{P(\boldsymbol{x}^{t})}[G(\boldsymbol{x}^{t})] \right\|_{\mathcal{H}_{\text{HDM}^*}^{2}} \\
			& \hspace*{-3.5em} + \left( \frac{2}{1 + \exp(-\chi \varepsilon)} - 1 \right) 
			\sum_{c=1}^{R} \biggl\| 
			\mathbb{E}_{P(\boldsymbol{x}^{s}|\boldsymbol{y}^{s}=c)}[G(\boldsymbol{x}^{s}) \mid \boldsymbol{y}^s=c ] \\
			& \hspace*{-3.1em} P(\boldsymbol{y}^{s}=c) 
			- \mathbb{E}_{P(\boldsymbol{x}^{t}|\boldsymbol{y}^{t}=c)}[G(\boldsymbol{x}^{t}) \mid \boldsymbol{y}^t=c] P(\boldsymbol{y}^{t}=c) 
			\biggr\|_{\mathcal{H}_{\text{HDM}^*}^{2}}
		\end{aligned}
		$}
\end{equation}

Computationally, evaluating $L_{\text{DDA}}$ over $N=n+m$ cases with $d$ kernels requires $\mathcal{O}((R+1)dN^2)\approx\mathcal{O}(RdN^2)$ time per iteration; therefore, the total time complexity over $T$ training iterations is $\mathcal{O}(RTdN^2)$.

\begin{figure}[t]
	\centering
	\includegraphics[width=0.49\textwidth]{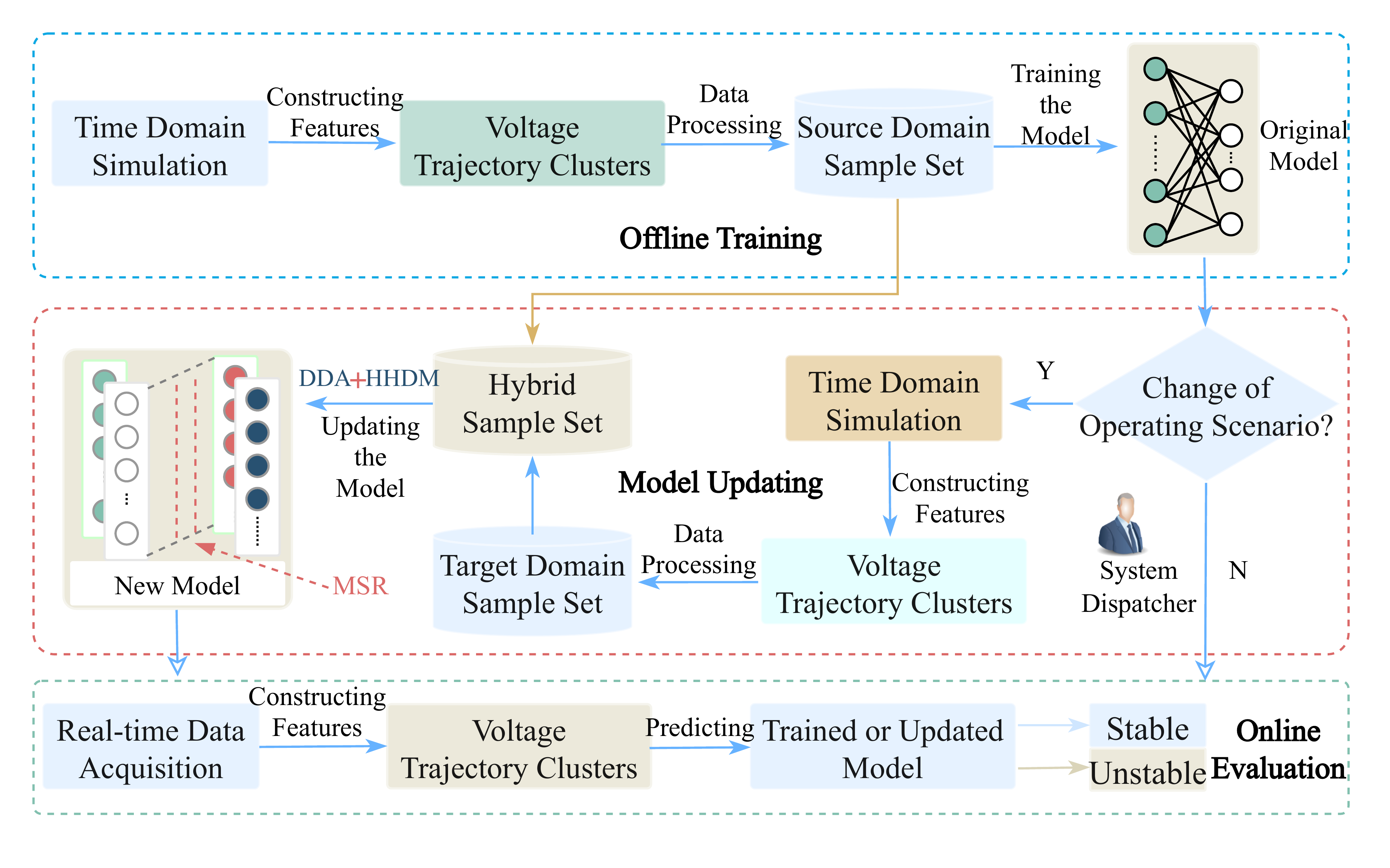}
	
	\caption{Adaptive TSA process.}
	\label{fig:Framework Process}
\end{figure}

\subsection{Loss Function and Parameter Update}

Considering both the prediction accuracy and generalization ability of the model, the objective function \( L \) of the HHDM-ATSA framework is defined as:
\begin{equation}
	L = L_{s} + L_{t} + \lambda L_{\mathrm{DDA}}
\end{equation}
where \( L_{s} \) and \( L_{t} \) denote the classification losses for the source and target domains, respectively, and \( \lambda \) is a weighting.

The loss terms \( L_{s} \), \( L_{t} \), and \( L_{\mathrm{DDA}} \) are jointly optimized to improve the assessment performance while facilitating the extraction of domain-invariant feature representations between the source and target domains. In each epoch of training, with learning rate $\gamma$, the parameters $\theta_G$ of the HHDM-ATSA are updated according to
\begin{equation}
	\theta_G \leftarrow \theta_G - \gamma \left( \frac{\partial L_s}{\partial \theta_G} + \frac{\partial L_t}{\partial \theta_G} + \lambda \frac{\partial L_{\text{DDA}}}{\partial \theta_G} \right)
\end{equation}

\section{Implementation of DA-Enabled TSA}

\subsection{Realization of Adaptive TSA}

The HHDM-ATSA framework includes three stages: offline training, model updating, and online evaluation. The adaptive TSA process is shown in Fig. 3.

\subsubsection{Offline Training} For a specific power system, considering various representative operating conditions and transient events, a raw case repository is first produced via TDS. In each event, an observation time window (OTW) with length $\Delta T$ and sampling rate $\Delta t$ is utilized to acquire $L$-point $(L = \Delta T / \Delta t)$ voltage profiles from individual buses. Here $T_{\text{win}}$ is set to begin from the steady-state time instant being closest to fault occurrence, and $\Delta T$ is set to be larger than the fault duration. By doing so, the acquired voltage profiles comprehensively covering pre-fault, fault-on, and post-fault system states would be quite informative for characterizing system transient responses after large disturbances. Afterwards, 29 trajectory cluster-based features are extracted from the acquired voltage profiles to construct the source-domain dataset. A 1D-CNN is then trained on this dataset to derive the initial TSA model.

\subsubsection{Model Updating} Upon detecting the variations of system operating conditions, a small number of target-domain cases are generated and labeled through TDS. These cases are combined with the source-domain dataset to form a hybrid dataset, which is then fed into the HHDM-ATSA framework to initiate the TSA model update process. Within this framework, the DDA strategy leverages the HHDM to quantify the discrepancies of both marginal and conditional probability distributions between the source and target domains. The TSA model’s parameters are then iteratively optimized to minimize these differences, thereby achieving fine-grained domain alignment and improving the TSA performance under the new scenarios. In the model update process, inspired by \cite{10634863}, the MSR algorithm is introduced to impose sparsity constraints on feature representation learning, so as to promote a more uniform distribution in the feature space. Let \( \bm{\mathcal{X}} \) denote the input feature matrix and \( \bm{\mathcal{Y}} \) represent the feature matrix after MSR application. The MSR is described by

\begin{equation}
	\bm{\mathcal{Y}}_1 = \bm{\mathcal{X}} / f(f(\bm{\mathcal{X}})) = \mathcal{X}_{ij} \bigg/ f\left( { \sum_{k=1}^{m} \mathcal{X}_{kj}^{2} + \varepsilon } \right)^{\tfrac{1}{p}}
\end{equation}

\begin{equation}
	\bm{\mathcal{Y}} = \bm{\mathcal{Y}}_1 / f(\bm{\mathcal{Y}}_1) = \mathcal{Y}_{ij} \bigg/ \left({ \sum_{k=1}^{m} \mathcal{Y}_{ik}^{2} + \varepsilon } \right)^{\tfrac{1}{p}}
\end{equation}
where $f = \left({ \sum_{k=1}^{m} \mathcal{X}_{kj}^{2} + \varepsilon } \right)^{\tfrac{1}{p}}$, $\varepsilon = 1\text{e-8}$.

\subsubsection{Online Evaluation} Real-time operating data of the grid are continuously collected, from which voltage trajectory cluster features are extracted and preprocessed. Such data are fed into the latest TSA model to rapidly issue online TSA~results.

\subsection{Performance Evaluation Indices}
The evaluation indices adopted in this paper include  $A_{\text{cc}}$,  $T_{\text{sr}}$, $T_{\text{ur}}$, and $G_{\text{mean}}$. These indices are defined as:

\begin{equation}
	A_{\text{cc}} = \frac{T_{\text{P}} + T_{\text{N}}}{T_{\text{P}} + T_{\text{N}} + F_{\text{P}} + F_{\text{N}}}
\end{equation}

\begin{equation}
	T_{\text{sr}} = \frac{T_{\text{P}}}{T_{\text{P}} + F_{\text{P}}}
\end{equation}

\begin{equation}
	T_{\text{ur}} = \frac{T_{\text{N}}}{F_{\text{N}} + T_{\text{N}}}
\end{equation}

\begin{equation}
	G_{\text{mean}} = \sqrt{T_{\text{sr}} \cdot T_{\text{ur}}}
\end{equation}
where \( T_{\text{P}} \) and \( F_{\text{P}} \) denote the correctly and incorrectly classified stable cases, respectively, while \( T_{\text{N}} \) and \( F_{\text{N}} \) represent the correctly and incorrectly classified unstable cases, respectively.

\section{Case study}
The efficacy of the HHDM-ATSA framework is comprehensively validated on the IEEE 39-bus system, the IEEE 162-bus system, and the Nordic test system. Numerical simulations are performed in batches using PSD-BPA, a commercial power system simulation package developed by China-EPRI. The DL models in the framework are implemented on the PaddlePaddle platform.

\subsection{ Case Studies on IEEE 39-bus System}
\subsubsection{Simulation Setup and Case Generation}
\begin{figure}[t]
	\centering
	\includegraphics[width=0.38\textwidth]{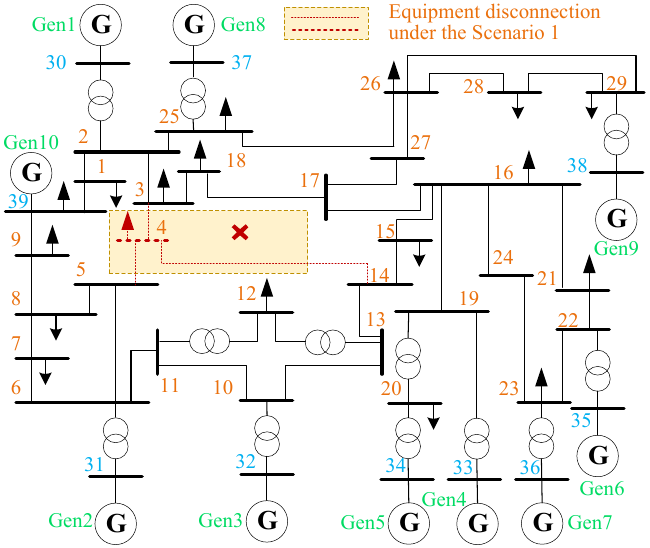}
	\caption{ One-line diagram of the IEEE 39-bus system.}
	\label{fig:IEEE39struct}
\end{figure}
In the IEEE 39-bus system, as illustrated in Fig. 4, let the base operating point be the original condition. To mimic variations of system operating scenarios, Scenario 1 considers an equipment failure where bus 4 and its three associated transmission lines are disconnected from the system. Scenario 2 models an overload loading condition by uniformly scaling all system loads to 130\% of the base level (while keeping the network topology unchanged). Generator outputs are redispatched to meet a new power flow balance, thereby forming a new condition that exhibits significant deviations from the original condition. Accordingly, three datasets are generated:

Dataset A (source domain): A wide variety of representative operating conditions and contingencies are considered here to cover diverse critical transient scenarios. Specifically, variations in operating conditions are generated by randomly setting diverse load power ranging from 80\%$\sim$120\% of the base levels, 10 typical three-phase short-circuit faults across each transmission line (with different fault locations), and shifts of fault duration times (ranging from 5 to 20 cycles). Based on these settings, 17,000 cases are generated by batch TDS, with unstable ones accounting for 46.7\%. Among the 17,000 cases, 13,000 ones are randomly selected for offline training, while the remaining 4,000 ones are reserved for online testing. In each case, an OTW with a length $\Delta T=0.5 $ s and a sampling interval of $\Delta t=0.01 $ s is utilized to acquire transient trajectory data ($L=50$ sampling points, $n_c=29$ features).

Dataset B (Target domain): In Scenario 1, the settings are kept the same as the source domain. 2,700 cases —1,480 stable and 1,220 unstable— are stochastically generated and then split into training and testing sets in a ratio of 1:2.

Dataset C (Target domain): In Scenario 2, 2700 cases are generated using the same settings as the source domain, with unstable cases accounting for 44.44\% of the dataset. The cases are divided into a training set for learning (900 cases) and a testing set for testing (1800 cases).

\subsubsection{Performance under Original Scenarios}
To illustrate the advantage of 1D-CNN, comparative tests are carried out here by considering other representative algorithms for TSA implementation, including decision tree (DT), support vector machine (SVM), random forest (RF), and k-nearest neighbor (KNN). For the 1D-CNN, its batch size is set to 150, and the learning rate is specified as 0.001. The online TSA performances of these models  are summarized in Table I. Evidently, none of these competitors defeats the 1D-CNN. Specifically, 1D-CNN achieves highly reliable stability prediction, with all of the four performance indices remaining above 98\%. In contrast, the data-driven algorithms, i.e., DT, SVM, RF, and KNN, exhibit larger prediction errors compared with the 1D-CNN, leading to a relatively low accuracy. Hence, 1D-CNN is taken as the base TSA model in subsequent case studies.

\begin{table}[t]
	\centering
	\caption{TSA Performance of Different Models}
	\renewcommand{\arraystretch}{1}
	\small 
	\begin{tabular}{
			>{\centering\arraybackslash}m{1.3cm}
			>{\centering\arraybackslash}m{1.3cm}
			>{\centering\arraybackslash}m{1.3cm}
			>{\centering\arraybackslash}m{1.3cm}
			>{\centering\arraybackslash}m{1.3cm}
			>{\centering\arraybackslash}m{1.3cm}
			>{\centering\arraybackslash}m{1.3cm}
		}
		
		\toprule
		Method & $A_{\text{cc}}$/\% & $T_{\text{sr}}$/\% & $T_{\text{ur}}$/\% & $G_{\text{mean}}$/\% \\
		\midrule
		1D-CNN & 98.23 & 98.13 & 98.32 & 98.22 \\
		DT     & 95.71 & 95.35 & 96.14 & 95.74 \\
		SVM    & 93.37 & 92.44 & 94.49 & 93.46 \\
		RF     & 95.01 & 94.63 & 95.34 & 94.98 \\
		KNN    & 93.69 & 92.67 & 94.90 & 93.78 \\
		\bottomrule
	\end{tabular}
\end{table}

\newcommand{\bnum}[1]{\textcolor{blue}{#1}}

\begin{table}[t]
	\centering
	\caption{Test results of original TSA model on Scenario 1 and 2}
	\label{tab:perf}
	\renewcommand{\arraystretch}{1}
	\small
	
	{
		\begin{tabular}{
				>{\centering\arraybackslash}m{1.1cm}
				>{\centering\arraybackslash}m{0.9cm}
				>{\centering\arraybackslash}m{1.0cm}
				>{\centering\arraybackslash}m{1.0cm}
				>{\centering\arraybackslash}m{1.0cm}
				>{\centering\arraybackslash}m{1.0cm}
			}
			\toprule
			Dataset & split & $A_{\text{cc}}$/\% & $T_{\text{sr}}$/\% & $T_{\text{ur}}$/\% & $G_{\text{mean}}$/\% \\
			\midrule
			\multirow{2}{*}{Dataset B} & Train & 89.03 & 87.37 & 90.59 & 88.97 \\
			& Test  & 88.87 & 86.97 & 90.77 & 88.84 \\
			\midrule
			\multirow{2}{*}{Dataset C} & Train & 89.87 & 88.55 & 91.15 & 89.84 \\
			& Test  & 89.51 & 89.47 & 89.52 & 89.49 \\
			\bottomrule
		\end{tabular}
	} 
	
\end{table}

\subsubsection{ Domain Adaptation in Unforeseen Scenarios}
In the presence of two new operating scenarios, each with 2,700 unforeseen cases each (i.e., datasets B and C from Scenarios 1 and 2, respectively), the model pre-trained under the original operating scenario (source domain) is first directly applied to each target domain without adaptation to examine its adaptability. As shown in Table II, the TSA accuracy drops to below 90\% in both Scenario 1 and 2, being unable to meet the practical demand for the reliability of online transient stability monitoring. Therefore, DA based on the proposed HHDM-ATSA framework is necessary. To this end, for target-domain datasets B and C, 500, 600, 700, 800, and 900 cases are randomly chosen from each dataset and integrated with the source-domain dataset to form ten hybrid case repositories, denoted as \( S_1\!\sim\!S_5 \) (from dataset B) and \( S_6\!\sim\!S_{10} \) (from dataset C), respectively. The HHDM-ATSA framework is configured with a batch size of 200, consisting of 170 source-domain cases and 30 target-domain cases. The weight parameters use in \( h_{\text{gs}}(\boldsymbol{x}^{s}, \boldsymbol{x}^{t}) \), specifically \( \alpha = 0.52 \), \( \beta = 0.48 \), \( [ \eta_1, \eta_2, \eta_3, \eta_4 ] = [0.20, 0.21, 0.02, 0.57] \), and \( [ \mu_1, \mu_2, \mu_3, \mu_4 ] = [0.33, 0.24, 0.14, 0.29] \), are optimized by the Optuna-based Bayesian optimization framework. The transfer learning performance of the HHDM-ATSA framework on these hybrid case repositories is presented in Fig. 5.

\begin{figure}[t]
	\centering
	\includegraphics[width=0.43\textwidth]{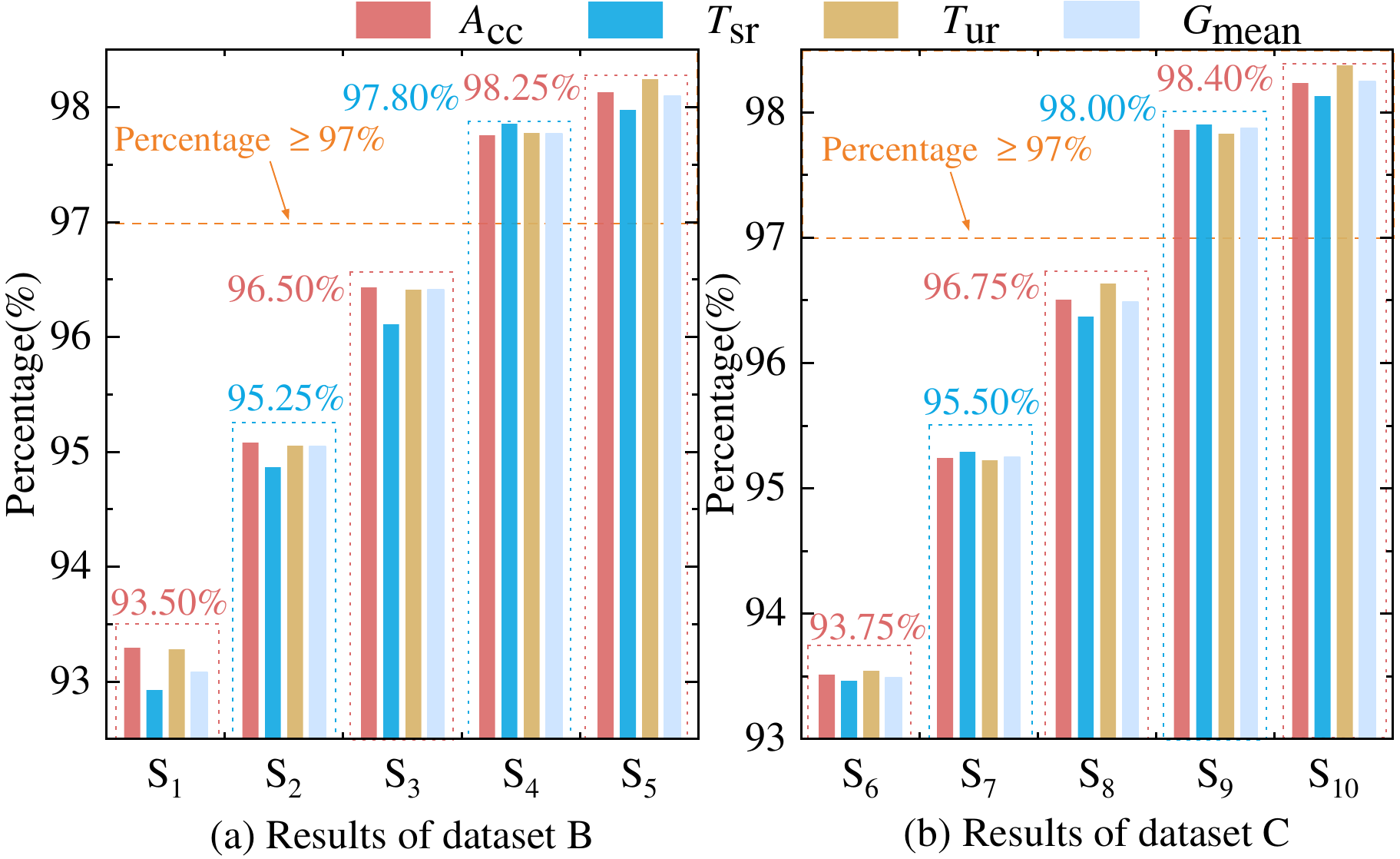}
	
	\caption{Transfer effect of the HHDM-ATSA framework.}
	\label{fig:39different-samples}
\end{figure}

\begin{figure}[t]
	\centering
	\includegraphics[width=0.43\textwidth]{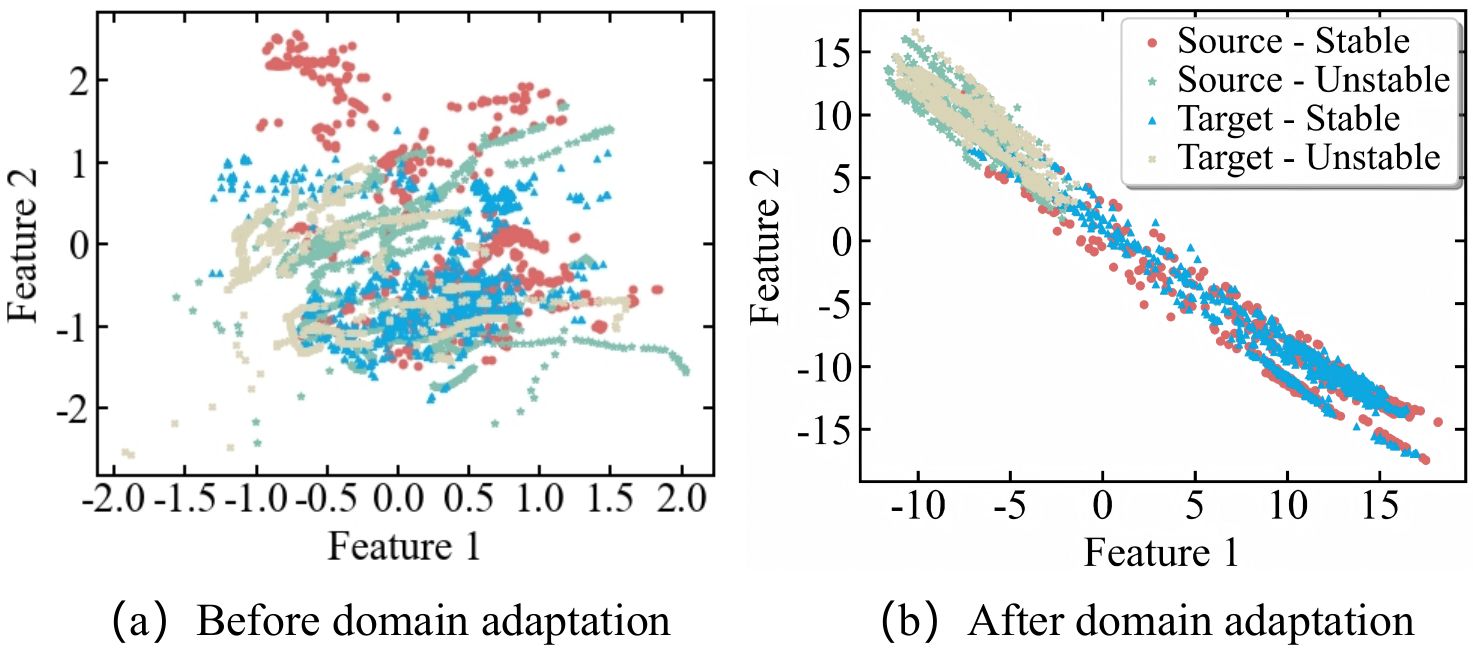}
	
	\caption{Comparison of data distributions before and after DA.}
	\label{fig:Fig6}
\end{figure}

As can be seen, for both scenarios, when the number of target-domain cases is fewer than 800 (i.e., $S_1\!\sim\!S_3$ and \( S_6\!\sim\! S_8 \)), the HHDM-ATSA framework has not yet achieved satisfactory TSA performance, with all four performance indices remaining below 96.75\%. With the increasing of cases, more domain-invariant features are extracted, leading to a gradual improvement of the overall prediction performance. In particular, when the case number reaches 800, the prediction accuracy is improved to 97.75\% and 97.86\%; with a further increase to 900, it reaches 98.13\% and 98.23\%. Notably, Scenario 1 exhibits a more pronounced domain shift compared to Scenario 2, and thus Scenario 1 is selected as the target domain in subsequent experiments. However, it should be mentioned that an excessively large number of cases would also introduce additional computational burdens in TDS and model updating, which is undesirable in practice. Therefore, considering the trade-off between prediction performance and model updating efficiency, the case number in the target domain is set to 800 ($S_4$) in subsequent tests. With this setting, the HHDM-ATSA framework would be able to achieve desirable DA performance with high computational efficiency.

Further, to intuitively illustrate the DA efficacy of the HHDM-ATSA, the distribution of cases within the extracted two-dimensional feature spaces is visualized here. Specifically, 2000 cases are chosen from the source and target domains, and are projected into the 1D-CNN-based feature spaces, as shown in Fig. 6. By comparison, the feature distributions of the two domains after DA are largely overlapped with each other, with distributional discrepancies significantly reduced. Moreover, it is observed that the feature space derived from DA is capable of linearly separating stable and unstable cases more purely. In fact, this also reveals that the HHDM-ATSA framework can capture more discerning domain-invariant stability features to better characterize critical transient stability features, thereby resulting in more reliable online TSA.

\subsubsection{Ablation Experiment}
To verify the effectiveness of each constituent module within the HHDM-ATSA framework, its three simplified versions are considered for ablation studies:

\textit{Scheme 1}: The MSR algorithm is removed, while all other configurations remain consistent with the HHDM-ATSA.

\textit{Scheme 2}: The Student kernel function is removed, and only the Gaussian kernel is adopted for DA, while all other configurations remain consistent with the HHDM-ATSA.

\textit{Scheme 3}: The Gaussian kernel function is removed, and only the Student kernel is adopted for DA, while all other configurations remain consistent with the HHDM-ATSA.

\textit{Scheme 4}: The CPDA mechanism is excluded. Only the marginal probability distributions between the source and target domains are aligned, with all other settings kept the same as in HHDM-ATSA.

\textit{Scheme 5}: The MPDA mechanism is discarded. Only the conditional probability distributions between the source and target domains are aligned, with the remaining settings identical to HHDM-ATSA.

\begin{table}[t]
	\centering
	\caption{Performance of Different Frameworks}
	\renewcommand{\arraystretch}{1}
	\small 
	\begin{tabular}{
			>{\centering\arraybackslash}m{1.3cm}
			>{\centering\arraybackslash}m{1.3cm}
			>{\centering\arraybackslash}m{1.3cm}
			>{\centering\arraybackslash}m{1.3cm}
			>{\centering\arraybackslash}m{1.3cm}
		}
		
		\toprule
		Scheme & $A_{\text{cc}}$/\% & $T_{\text{sr}}$/\% & $T_{\text{ur}}$/\% & $G_{\text{mean}}$/\% \\
		\midrule
		Proposed  & 97.75  & 97.77  & 97.85  & 97.78  \\
		Scheme 1    & 97.53 & 97.68 & 97.40 &97.55 \\
		Scheme 2 & 97.38  & 97.43  & 97.30  & 97.36  \\
        Scheme 3 & 97.35 & 99.32 & 97.74 & 97.52 \\
		Scheme 4     & 97.24 & 97.21 & 97.28 & 97.25 \\
		Scheme 5    & 97.14 & 96.51 & 97.84 & 97.17 \\
		\bottomrule
	\end{tabular}
\end{table}

Table III summarizes the five schemes' online TSA performance after DA. Obviously, the five simplified variations perform worse than the HHDM-ATSA framework, with their prediction accuracies being less than 97.6\%. The superior performance of HHDM-ATSA can be mainly attributed to its special attention to long-tail patterns of transient trajectory data, precise alignment at the sub-domain level,  and more uniform feature learning via the MSR mechanism, which can sufficiently learn critical domain-invariant transient stability features during model updating. Consequently, the proposed framework facilitates a more rapid and effective TSA model adaptation under dynamically varying system conditions.

\begin{figure}[t]
	\centering
	\includegraphics[width=0.40\textwidth]{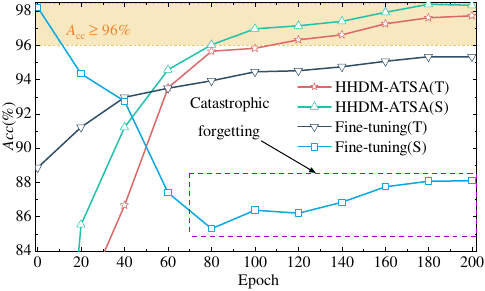}
	\caption{Verification of resistance to catastrophic forgetting.}
	\label{fig:Sustainable-Learning}
\end{figure}

\subsubsection{Verification of Resistance to Catastrophic Forgetting}
Considering the fact that frequent changes of power system operating conditions are unavoidable in practice, the proposed approach’s capability in resisting catastrophic forgetting is further tested here. Specifically, here a fine-tuning algorithm is considered for comparison, with the target-domain cases from $S_4$ fed into the pre-trained model to fine-tune the entire network parameters. The TSA results of the HHDM-ATSA framework and this fine-tuning alternative on testing cases are shown in Fig. 7. Here HHDM-ATSA(T) and HHDM-ATSA(S) denote the accuracy curves of HHDM-ATSA on the target- and source-domain test sets, respectively, while Fine-tuning(T) and Fine-tuning(S) represent those of the fine-tuning algorithm.

It is found that, as training progresses, the fine-tuning algorithm gradually improves its prediction accuracy in the target domain. However, it simultaneously suffers from persistent performance degradation in the source domain. In contrast, after DA, the proposed HHDM-ATSA framework achieves an online TSA accuracy of more than 97.5\% in the source and target domains.  The results verify that the framework’s strong capability in resistance to catastrophic forgetting and accumulation of new stability knowledge from new operating conditions for reliable online TSA, without forgetting old knowledge initially learned in the source domain.

\subsubsection{Comparative Study}
To comprehensively verify the advantage of the proposed HHDM-ATSA framework, more extensive tests are performed by comparing it with several well-known transfer learning schemes:

\textit{Scheme 1}: No transfer learning is employed. Only the target-domain cases from set $S_4$ are used for training, and a new model is trained from scratch.

\textit{Scheme 2}: An instance-based transfer learning strategy is adopted,  in which set $S_4$ is the training set to train the model.

\textit{Scheme 3}: A parameter-based transfer learning approach is applied, where the architecture and parameters of the source model are transferred to the target domain. Then, $S_4$ is utilized to fine-tune the entire model.

\textit{Scheme 4}: A hybrid transfer learning strategy that integrates model parameter-based and instance-based transfer. The source model’s architecture and parameters are transferred to the target domain, and the set $S_4$ is used for fine-tuning.

\textit{Scheme 5}: A supervised domain adversarial transfer learning method is adopted, incorporating a gradient reversal layer to adversarially train the domain discriminator (domain-level adaptation only). The training set is $S_4$.

\textit{Scheme 6}: A domain-conditioned joint adaptation strategy is employed \cite{10017183}, where domain- and class-level adaptation are jointly achieved via domain-adversarial training and bi-classifier adversarial learning. The model is learned using $S_4$.

\textit{Scheme 7}: A dynamic hybrid domain adaptation method is adopted \cite{LU2025104172}, in which domain-level adaptation is performed via MMD, while class-level adaptation is enforced through domain-adversarial learning. The training set is $S_4$.

\textit{Scheme 8}: A low-rank sparse generative adversarial domain adaptation method is employed \cite{10193868}, in which a spectral low-rank dictionary learning technique for feature sparsification, and joint domain-adversarial training are used to align the global and class-level distributions across domains. The training set is $S_4$.

\textit{Scheme 9}: A joint distribution adaptation approach is employed \cite{10251665}, where the MMD is incorporated to align the marginal and class-conditional distributions between the source and target domains, and a balance factor is further introduced to adaptively weight the two alignment terms. $S_4$ serves as the training set.

\textit{Scheme 10}: Beyond MMD-based marginal and class-conditional probability distribution alignment, intra-class distances are further minimized while inter-class distances are maximized, thereby improving the discriminability of the DA framework \cite{rezaei2024visual}. The set $S_4$ is adopted for training.

\textit{Proposed Scheme}: The proposed HHDM-ATSA framework, with $S_4$  serving as the training set.

\begin{table}[t]
	\centering
	
	\caption{Effect comparison of Different transfer Schemes} 
	\renewcommand{\arraystretch}{1} 
	\small 
	\begin{tabular}{
			>{\centering\arraybackslash}m{1.5cm}
			>{\centering\arraybackslash}m{1.1cm}
			>{\centering\arraybackslash}m{1.3cm}
			>{\centering\arraybackslash}m{1.3cm}
			>{\centering\arraybackslash}m{1.3cm}
		}
		\toprule
		Scheme & $A_{\text{cc}}$/\% & $T_{\text{sr}}$/\% & $T_{\text{ur}}$/\% & $G_{\text{mean}}$/\% \\
		\midrule
		Proposed& 97.75  & 97.77  & 97.85  & 97.78  \\
		\hline
		Scheme 1 & 93.26 & 92.97 & 93.50 & 93.24 \\
		Scheme 2 & 95.12 & 95.05 & 95.19 & 95.12 \\
		Scheme 3 & 95.36 & 95.57 & 95.18 & 95.38 \\
		Scheme 4 & 96.10 & 96.60 & 95.67 & 96.13 \\
		Scheme 5 & 96.32 & 96.49 & 96.11 & 96.29 \\
		Scheme 6 & 96.74 & 97.13 & 96.40 & 96.76 \\
		Scheme 7 & 96.95 & 96.84 & 97.01 & 96.92 \\
		Scheme 8 & 97.00    & 97.11 & 96.82 & 96.96 \\
		Scheme 9 & 97.10 & 97.14 & 97.12 & 97.12 \\
		Scheme 10 & 97.29 & 97.20 & 97.35 & 97.26 \\
		\bottomrule
	\end{tabular}
	
	\arrayrulecolor{black} 
\end{table}

As shown in Table IV, Scheme 1, trained from scratch with insufficient cases, fails to effectively capture the intricate stability features in new scenarios, leading to an accuracy below 95\%. With further attention to Schemes $2\!\sim\!4$, it can be seen that,  compared with the two simplified DA versions, i.e., Schemes 2 and 3, Scheme 4 is able to improve the TSA accuracy by 0.98\% and 0.79\%. Scheme 5 adopts domain-adversarial learning for marginal distribution alignment and achieves a TSA accuracy of 96.32\%.  Additionally, to enable more fine-grained DA, Schemes 6$\sim$10 jointly consider both global and class-level adaptation, implemented via domain-adversarial learning and/or MMD-based discrepancy minimization to reduce the distribution shift across domains.  Schemes 6, 7, and 8 perform slightly inferior to Schemes 9 and 10, mainly owing to the training instability of adversarial optimization.  Notably, Scheme 10 further improves the adaptation by compacting intra-class samples and enlarging inter-class separations on top of marginal and class-conditional alignment, yielding outperforming Schemes 6, 7, 8, and 9 by 0.55\%, 0.34\%, 0.29\%, and 0.19\%, respectively. Compared with Schemes $5\mathord{\sim}10$, the proposed HHDM-ATSA framework demonstrates better online TSA performance, being able to accurately identify 0.5\%$\sim$1.5\% more of transient cases. By further comparing their technical differences, it is found that Scheme 5 does not align sub-domain–level distributions, while Scheme 6$\sim$10 fails to efficiently handle the long-tail features in transient stability data. Moreover, Schemes 9 and 10 align class-conditional distributions as a surrogate for conditional alignment, and thus do not achieve a strict alignment of the conditional probability distribution. This implies that the proposed framework with special attention to long-tail patterns of transient trajectory data and refined-level distribution alignment is capable of making the TSA model adapt well to unforeseen scenarios during online application.

To further evaluate the overall learning efficiency of the HHDM-ATSA, its time complexity for updating the TSA model under new operating scenarios is tested. With the target of remaining the TSA accuracy above 96\%, four metrics, including the required minimum number of target-domain cases , TDS time, model training time, and total runtime, are considered for learning efficiency evaluation. In addition to the HHDM-ATSA, the total learning efficiencies of the 10 comparative schemes involved above are also examined here for comparison. The results are shown in Table V.

\begin{table}[t]
	\centering
	\caption{Efficiency of Different transfer Schemes}
	\renewcommand{\arraystretch}{1}
	\small 
	\begin{tabular}{
			>{\centering\arraybackslash}m{1.5cm}
			>{\centering\arraybackslash}m{1.1cm}
			>{\centering\arraybackslash}m{1.3cm}
			>{\centering\arraybackslash}m{1.3cm}
			>{\centering\arraybackslash}m{1.3cm}
			>{\centering\arraybackslash}m{1.3cm}
		}
		
		\toprule
		Scheme & RTDC & TDST (s) & MTT (s) &  TRT (s) \\
		\midrule
		Proposed& 645 & 2240 & 214 & 2454 \\
		\hline
		Scheme 1 & 8600 & 29867 & 78 & 29945 \\
		Scheme 2 & 1100 & 3850 &  105 & 3955 \\
		Scheme 3 & 1030 & 3605 & 33  & 3638 \\
		Scheme 4 & 790 & 2765 & 110 & 2875 \\
		Scheme 5 & 765 & 2679 & 85 & 2764 \\
		Scheme 6 & 720 & 2520 & 127 & 2647 \\
		Scheme 7 & 710 & 2485 & 187 & 2672 \\
		Scheme 8 & 710 & 2485 & 179 & 2664 \\
		Scheme 9 & 700 & 2450 & 198 & 2648 \\
		Scheme 10 & 685 & 2397 & 235 & 2632 \\
		\bottomrule
	\end{tabular}
	\vspace{0.3em}
	
	\raggedright \textit{\textbf{ Note 1:}} The 2nd to 5th columns represent the required minimum number of target-domain cases (RTDC), TDS time consumption (TDST), model training time (MTT), and total runtime (TRT = TDST + MTT). To make a fair comparison, all the schemes are built with the same learning target of maintaining the overall TSA accuracy at the consistent level of \(\geq\)96\%.
\end{table}

As can be observed, the proposed HHDM-ATSA framework is able to complete the procedures of TDS and model training within 2300 s and 220 s, respectively. Although its training efficiency is not the highest among the compared methods, it does not essentially impair its online application efficiency. More importantly, it is found that the proposed framework achieves the shortest total runtime, implementing the whole learning procedure with 3$\sim$458 mins faster than the other alternatives. In fact, with the overall runtime mainly cost by TDS, the difference in total runtime consumption between different approaches is primarily attributed to the number of cases required for TDS. Benefiting from the requirement of the fewest target-domain cases for TDS, the proposed framework thus defeats others regarding the overall learning efficiency. Such superior DA efficiency is of great significance for practical TSA applications.

\begin{table}[!t]
	\centering
	\caption{Dataset Statistics on the IEEE 162-Bus System}
	\label{tab:dataset_stats}
	
	\renewcommand{\arraystretch}{1}
	\small
	
	\newcommand{\thickhline}{%
		\noalign{\global\arrayrulewidth=0.8pt}\hline
		\noalign{\global\arrayrulewidth=0.4pt}}
	
	{
		\begin{tabular}{
				>{\raggedright\arraybackslash}m{1.6cm}
				>{\centering\arraybackslash}m{1.3cm}
				>{\centering\arraybackslash}m{1.3cm}
				>{\centering\arraybackslash}m{1.3cm}
				>{\centering\arraybackslash}m{1.3cm}
			}
			\thickhline
			\multirow{2}{*}{Dataset} & \multirow{2}{*}{Split} & \multicolumn{2}{c}{Labels} & \multirow{2}{*}{Total} \\
			\cline{3-4}
			&                & \text{Stable} & \text{Unstable} & \\
			\hline
			\multirow{2}{*}{Dataset D}
			& Train & 8250 & 6930 & 15180 \\
			& Test  &  2147 & 2073&  4220  \\
			\hline
			\multirow{2}{*}{Dataset E}
			& Train & 574 & 426& 1000  \\
			& Test   &  2125 & 1700& 3825 \\
			\thickhline
		\end{tabular}
	} 

\end{table}

\begin{table}[t]
	\centering
	\caption{Test Results on the IEEE 162-Bus System}
	\label{tab:162-result}
	
	\renewcommand{\arraystretch}{1}
	\small
	
	\newcommand{\thickhline}{%
		\noalign{\global\arrayrulewidth=0.8pt}\hline
		\noalign{\global\arrayrulewidth=0.4pt}}
	
	{
		\begin{tabular}{
				>{\centering\arraybackslash}m{1.5cm}
				>{\centering\arraybackslash}m{0.8cm}
				>{\centering\arraybackslash}m{1.0cm}
				>{\centering\arraybackslash}m{1.0cm}
				>{\centering\arraybackslash}m{1.0cm}
				>{\centering\arraybackslash}m{1.0cm}
			}
			\thickhline
			Scheme & Set  & $A_{\text{cc}}$/\% & $T_{\text{sr}}$/\% & $T_{\text{ur}}$/\% & $G_{\text{mean}}$/\% \\
			\hline
			
			\multirow{2}{*}{1D-CNN}
			& Test(D) & 98.60 & 98.26 & 98.93 & 98.59 \\
			& Test(E)  & 83.93 & 82.75 & 85.06 & 83.90 \\
			\hline
			
			Proposed & Test(E) & 98.48 & 98.41 & 98.53 & 98.47 \\
			\hline
			
			Scheme 5 & Test(E) & 96.62 & 96.20 & 97.21 & 96.70 \\
			Scheme 6 & Test(E) & 97.13 & 96.91 & 97.33 & 97.12 \\
			Scheme 7 & Test(E) & 97.20 & 97.07 & 97.36 & 97.21 \\
			Scheme 8 & Test(E) & 97.38 & 97.14 & 97.59 & 97.36 \\
			Scheme 9 & Test(E) & 97.48 & 97.27 & 97.74 & 97.50 \\
			Scheme 10 & Test(E) & 97.81 & 97.67 & 97.93 & 97.87 \\
			\thickhline
		\end{tabular}
	} 

\end{table}

\subsection{Case Studies on IEEE 162-bus System}
\subsubsection{System Description and Data Generation}
To analyze the scalability of the HHDM-ATSA framework under complex network structures and complicated operating conditions, the IEEE 162-bus system is adopted for additional tests. The system consists of 43 generators, 95 buses, and 238 transmission lines. Following a similar simulation setup as in the IEEE 39-bus system, various representative operating scenarios, load power, fault types,  fault locations, and fault duration times are taken into comprehensive account for TSA case generation. Leveraging batch numerical simulations in PSD-BPA, two datasets are generated: dataset D corresponding to the original topology (source domain) and dataset E representing $N-3$ topological modifications (target domain). The generated datasets are statistically summarized in Table VI.

\subsubsection{DA Performance Analysis}
The DA performance of the HHDM-ATSA framework is analyzed under distribution shifts from the source domain to the target domain induced by topology variations. Initially, a 1D-CNN is trained only via the source domain's training set to construct a base TSA model. The resulting model is then directly deployed to the target domain for evaluation without any adaptation. Subsequently, besides the HHDM-ATSA, the DA performance of six comparative methods, namely Schemes 5$\sim$10 as detailed in Section VI-A.6), is evaluated, and the results are summarized in Table VII. When the base TSA model is directly applied to the target domain, its predictive performance deteriorates significantly. This performance drop indicates a significant mismatch between the source and target distributions induced by topology variations, rendering the direct deployment of the pre-trained model in the target domain unreliable. After applying DA, Schemes 5$\sim$10 mitigate the mismatch to varying extents, yielding evident performance recovery; however, none of the four evaluation metrics exceeds 98\%. In contrast, the HHDM-ATSA achieves the most desirable performance on the target domain, attaining an accuracy of 98.48\%. These results further demonstrate the advantage of HHDM-ATSA in effectively mitigating topology-induced domain shifts and improving the generalization capability of the TSA model.

\begin{table}[!t]
	\centering
	\caption{Dataset Statistics on the Nordic Test System}
	\label{tab:dataset_stats1}
	
	\renewcommand{\arraystretch}{1}
	\small
	
	\newcommand{\thickhline}{%
		\noalign{\global\arrayrulewidth=0.8pt}\hline
		\noalign{\global\arrayrulewidth=0.4pt}}
	
	{
		\begin{tabular}{
				>{\raggedright\arraybackslash}m{1.6cm}
				>{\centering\arraybackslash}m{1.3cm}
				>{\centering\arraybackslash}m{1.3cm}
				>{\centering\arraybackslash}m{1.3cm}
				>{\centering\arraybackslash}m{1.3cm}
			}
			\thickhline
			\multirow{2}{*}{Dataset} & \multirow{2}{*}{Split} & \multicolumn{2}{c}{Labels} & \multirow{2}{*}{Total} \\
			\cline{3-4}
			&                & \text{Stable} & \text{Unstable} & \\
			\hline
			\multirow{2}{*}{Dataset F}
			& Train & 8937 & 7334 & 16271  \\
			& Test  & 1820 & 1742 &  3562  \\
			\hline
			\multirow{2}{*}{Dataset G}
			& Train  & 558 & 442& 1000 \\
			& Test  &  1732 & 1546  &  3278\\
			\thickhline
		\end{tabular}
	} 
\end{table}

\begin{table}[t]
	\centering
	\caption{Test Results on the Nordic Test System}
	\label{tab:nordic-result}
	
	\renewcommand{\arraystretch}{1}
	\small
	
	\newcommand{\thickhline}{%
		\noalign{\global\arrayrulewidth=0.8pt}\hline
		\noalign{\global\arrayrulewidth=0.4pt}}
	
	{
		\begin{tabular}{
				>{\centering\arraybackslash}m{1.5cm}
				>{\centering\arraybackslash}m{0.8cm}
				>{\centering\arraybackslash}m{1.0cm}
				>{\centering\arraybackslash}m{1.0cm}
				>{\centering\arraybackslash}m{1.0cm}
				>{\centering\arraybackslash}m{1.0cm}
			}
			\thickhline
			Scheme & Set  & $A_{\text{cc}}$/\% & $T_{\text{sr}}$/\% & $T_{\text{ur}}$/\% & $G_{\text{mean}}$/\% \\
			\hline
			
			\multirow{2}{*}{1D-CNN}
			& Test(F) & 98.02 & 98.12 & 97.96 & 98.03 \\
			& Test(G)  & 81.91 & 85.96 & 78.33 & 82.06 \\
			\hline
			
			Proposed & Test(G) & 97.80 & 97.77 & 97.85 & 97.78 \\
			\hline
			
			Scheme 5 & Test(G) & 95.68 & 95.56 & 95.77 & 95.66 \\
			Scheme 6 & Test(G) & 96.17 & 96.11 & 96.24 & 96.18 \\
			Scheme 7 & Test(G) & 96.51 & 96.24 & 96.72 & 96.48 \\
			Scheme 8 & Test(G) & 96.50 & 96.11 & 96.85 & 96.49 \\
			Scheme 9 & Test(G) & 96.89 & 96.93 & 96.86 & 96.89 \\
			Scheme 10 & Test(G) & 97.16 & 97.18 & 97.13 & 97.15 \\
			\thickhline
		\end{tabular}
	} 
\end{table}

\subsection{Case Studies on Nordic Test System}
To better approximate practical operating conditions, the Nordic test system, modified from the actual Swedish and Nordic power grid, is adopted here for high-fidelity numerical simulations. The system consists of 23 generators, 41 buses, and 69 transmission lines. The transient case simulation settings remain consistent with those in Section VI-B.1), and two datasets are generated via TDS: Dataset F corresponding to the original topology (source domain) and Dataset G  corresponding to the $N-2$ topology (target domain). The generated datasets are statistically summarized in Table VIII. Moreover, the overall DA procedure and the settings of the comparison schemes follow those detailed in Section VI-B.2).

As shown in Table IX, the HHDM-ATSA framework achieves the best  DA performances among all methods. Specifically, its overall TSA classification accuracy is 0.7$\sim$2.5\% higher than that of the alternatives, which either neglect the long-tail features inherent in transient stability data or fail to implement a more fine-grained DA mechanism—both of which are effectively handled by the proposed. These superior performances imply that HHDM-ATSA can effectively perform DA tasks in the presence of diverse and severe domain shifts. Therefore, it can reliably serve as a model updating tool, rapidly improving online TSA performance under various changing or unforeseen scenarios in real-world operation.

\section{Conclusion}
To address the insufficient adaptability of existing data-driven TSA schemes in the presence of operational variations during online application, this work develops an intelligent HHDM-ATSA framework that enables a TSA model initially trained offline to adapt well to diverse new online operating conditions in a reliable and efficient way. By carefully designing an HHDM metric as well as a DDA strategy and incorporating them with the powerful MSR algorithm, the HHDM-ATSA framework is able to sufficiently learn the inherent domain-invariant transient stability features across diverse operating conditions. Extensive numerical tests on three benchmark test systems showcase highly competitive DA performance of the proposed scheme. Compared with existing solutions, it robustly achieves superior online TSA accuracy of more than 97.5\% under various unforeseen scenarios. Besides, it has a higher TSA model updating efficiency for online monitoring, with the need for less target-domain labeled cases and lighter computational burdens.

This study mainly considers exploiting the fewest labeled cases for the implementation of model online updating. Future work will be devoted to investigating unsupervised domain adaptation strategies to implement TSA-oriented DA in a cost-effective manner, without reliance on costly efforts for labeling cases generated in new scenarios. Meanwhile, how to address the reality gap between simulated environments and practical measurement conditions is planned as another research direction.

\bibliographystyle{IEEEtran}
\bibliography{TII-26-2461.R1}
\end{document}